\documentclass[twocolumn]{aastex63}
\usepackage{xcolor}
\usepackage{apjfonts}

\newcommand{\cacnchnh}{Ca-CN-CH-NH}

\newcommand{\nhjwl}{$nh_{\rm JWL}$}
\newcommand{\chjwl}{$ch_{\rm JWL}$}
\newcommand{\cnjwl}{$cn_{\rm JWL}$}

\newcommand{\cnw}{CN-w}
\newcommand{\cns}{CN-s}
\newcommand{\cnsi}{CN-s$_{\rm I}$}
\newcommand{\cnse}{CN-s$_{\rm E}$}

\newcommand{\hkjwl}{$hk_{\rm JWL}$}

\newcommand{\nrgbthree}{$n$(\cnw):$n$(\cnsi):$n$(\cnse)}
\newcommand{\nrgbsub}{$n$(\cnsi):$n$(\cnse)}

\newcommand{\pchjwl}{$\parallel ch_{\rm JWL}$}

\newcommand{\str}{Str\"omgren}

\newcommand{\vvhbmag}{$-$2 mag $\leq$ $V-V_{\rm HB}$ $\leq$ 2 mag}
\newcommand{\dy}{$\Delta$Y}

\newcommand{\cfe}{[C/Fe]}
\newcommand{\nfe}{[N/Fe]}
\newcommand{\feh}{[Fe/H]}
\newcommand{\scfe}{$\sigma$[C/Fe]}

\newcommand{\shk}{$\sigma$(hk)}

\newcommand{\cnwave}{$\lambda$3883}
\newcommand{\chwave}{$\lambda$4250}
\newcommand{\nhwave}{$\lambda$3360}
\newcommand{\masyr}{mas~yr$^{-1}$}

\shorttitle{M5}
\shortauthors{Lee}
\begin{document}

\title{Formation of Multiple Populations of M5 (NGC~5904)}

\author[0000-0002-2122-3030]{Jae-Woo Lee}
\affiliation{Department of Physics and Astronomy, Sejong University\\
209 Neungdong-ro, Gwangjin-Gu, Seoul, 05006, Republic of Korea\\
jaewoolee@sejong.ac.kr, jaewoolee@sejong.edu}

\begin{abstract}
With our new \cacnchnh\ photometry, we revisit the globular cluster (GC) M5. We find that M5 is a mono-metallic GC with a small metallicity dispersion. Our carbon abundances show that the \scfe\ of the M5 \cns\ population, with depleted carbon and enhanced nitrogen abundances, is significantly large for a single stellar population. Our new analysis reveals that the M5 \cns\ population is well described by the two stellar populations: the \cnsi, being the major \cns\ component, with the intermediate carbon and nitrogen abundance and the \cnse\ with the most carbon-poor and nitrogen-rich abundance. We find that the \cnse\ is significantly more centrally concentrated than the others, while \cnw\ and \cnsi\ have similar cumulative radial distributions. The red giant branch bump $V$ magnitude, the helium abundance barometer in mono-metallic populations, of  individual populations appears to be correlated with their mean carbon abundance, indicating that carbon abundances are anticorrelated with helium abundances. We propose that the \cnse\ formed out of gas that experienced proton-capture processes at high temperatures in the innermost region of the proto-GC of M5 that resided in a dense ambient density environment. Shortly after, the \cnsi\ formed out of gas diluted from the pristine gas in the more spatially extended region, consistent with the current development of numerical simulations by others.
\end{abstract}


\section{Introduction}
Recent photometric and spectroscopic studies of globular clusters (GCs) in our Galaxy and nearby galaxies revealed the ubiqutous nature of multiple populations (MPs) in GCs and opened a new golden era in the field of stellar populations \citep[e.g,][] {carretta09, lee09, milone17,  bastian18, gratton19}.
Previously considered as simple and dull, understanding the formation and evolution of GCs requires complex and somewhat fine-tuned processes.
One of the remarkable aspects is that the second generation (SG) of stars, which is believed to form in more spatially concentrated inner regions of GCs out of gas ejected from the clusters' first generation (FG) of stars \citep[e.g.,][]{dercole08}, is the major component of GCs with MPs.
Recently, sophisticated numerical simulations have become available and they help us to delineate how the second generation of stars formed and dynamically evolved under certain assumptions \citep[e.g.,][]{bekki19, calura19}.

Major observational breakthroughs have been made via the Hubble Space Telescope \citep[e.g.,][]{milone17} and the ground-based multi-object spectroscopy with large aperture telescopes \citep[e.g.,][]{carretta09}. However, these approaches have some drawbacks: broadband photometry has a potential degeneracy problem for individual elemental abundances, while conventional spectroscopy has a potential contamination problem by nearby stars in crowded regions.

Surface carbon and nitrogen abundances, the key elements in the study of GC MPs, can illuminate the internal nucleosynthesis of evolved GC stars and even an entire cluster's evolutionary history. However, due to observational limitations, both in the HST photometry and the ground-based spectroscopy that we mentioned above, reliable carbon and nitrogen abundances are often poorly known.

In this Letter, we revisit the GC M5 (NGC~5904) with our new photometric system, optimized to study carbon and nitrogen abundances of red giant branch (RGB) stars in crowded fields. In our previous Ca-CN photometric study of the cluster, we found that M5 contains two MPs with identical radial distributions but different structural and kinematical properties \citep{lee17,lee19a,lee19b}. 
Our new \cacnchnh\ observations of the cluster reveal new population with interesting physical properties that we were not able to detect in our previous study.
Here we present the \feh, \cfe, and \nfe\ of individual RGB stars and new insight on the formation and evolution of MPs in M5.

\section{New Observations}
During the last decade, we developed a new photometric system that can measure \ion{Ca}{2} H and K lines, NH, CN, and CH molecular band absorption strengths at \nhwave, \cnwave, and \chwave, respectively, to provide reliable \feh, \cfe, and \nfe\ of individual RGB stars using theoretical fine model grids with various stellar input parameters and elemental abundances \citep[see][for our new filter system]{lee15, lee17, lee19a, lee19b, lee21}. Our method can avoid the crowding effects with potential contamination by neighboring stars, which is one of the limitations in the classical spectroscopic study of the GC stars, in particular, in the central part of GCs. Therefore, our approach can guarantee a more complete sample for the GC study.

Observations for our \str\ and Ca-CN photometry were conducted using the CTIO 1.0 m telescope in 21 nights in seven runs from 2007 May to 2014 May \citep[see][for details]{lee17}. In addition, we  obtained the JWL34 and JWL43 photometry using the KPNO 0.9 m telescope in 11 nights in three separate runs from 2017 February and 2019 July.\footnote{The transmission functions for our JWL33 and JWL43 filters can be found in Figure~3 of \citet{lee21} and Figure~1 of \citet{lee19b}, respectively.}

The raw data handling were described in detail in our previous works \citep{lee15, lee17}. The photometry of M5 and standard stars were analyzed using DAOPHOTII, DAOGROW, ALLSTAR and ALLFRAME, and COLLECT-CCDAVE-NEWTRIAL packages \citep{pbs94}.
The total number of stars in our M5 field from our ALLFRAME run was more than 60,000.

Finally, we derived the astrometric solutions for individual stars using coordinates of more than 3500 stars extracted from the Gaia Early Data Release 3 \citep[EDR3,][]{gaiaedr3}  and the IRAF IMCOORS package. The rms errors of our fit are very small, 0\farcs008 and 0\farcs006 along the R.A.\ and the decl., respectively.

\begin{figure}
\epsscale{1.}
\figurenum{1}
\plotone{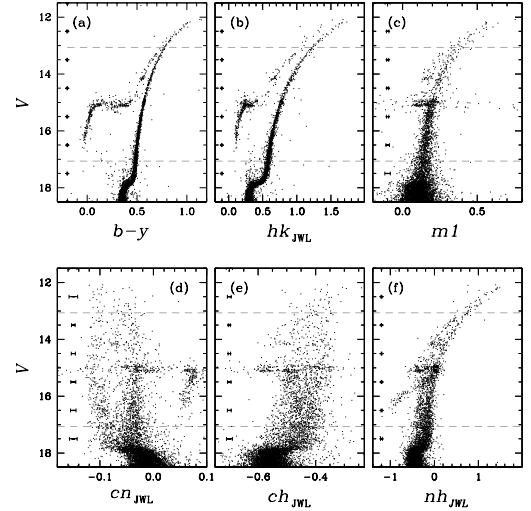}
\caption{
CMDs of M5 membership stars based on the proper motion study of the Gaia EDR3 \citep{gaiaedr3}. We only show the stars with radial distance from the center larger than 1\arcmin. Discrete bimodal RGB sequences can be seen in the M5 \cnjwl\ CMD, while a broad RGB sequence can be seen in the \chjwl,\nhjwl, and $m1$ CMDs \citep[see also][]{lee17}. 
}\label{fig:cmd}
\end{figure}

\section{Results}
\subsection{Color-Magnitude Diagrams}
Throughout this work, we will use our own photometric indices already defined in our previous works \citep{lee17, lee19a, lee19b, lee21},
\begin{eqnarray}
hk_{\rm JWL} &=& ({\rm Ca}_{\rm JWL} - b) - (b-y). \label{eq:hk} \\
nh_{\rm JWL} &=& ({\rm JWL34} - b) - (b-y). \label{eq:nh} \\
cn_{\rm JWL} &=& {\rm JWL39} - {\rm Ca}_{\rm JWL}, \label{eq:cn} \\
ch_{\rm JWL} &=& ({\rm JWL43} - b) - (b-y). \label{eq:ch} 
\end{eqnarray}
Note that the \hkjwl\ index measures the absorption strengths of \ion{Ca}{2} H and K lines and it is a good measure of metallicity \citep{att91,lee09,lee15}. The \nhjwl, \cnjwl, and \chjwl\ are excellent measures of the NH band at \nhwave, the CN band at \cnwave, and the CH G band at \chwave, respectively, for cool stars \citep{lee17,lee19b,lee20,lee21}.

We made use of the proper motions from the Gaia EDR3 to select the cluster's membership stars, following the method similar to those used in our previous studies \citep[see, e.g.,][and references therein]{lee20,lee21}.
We derived the mean values of proper motions of M5 with iterative sigma-clipping calculations, finding that, in units of \masyr, ($\mu_{\rm R.A.}\times\cos\delta$, $\mu_{\rm decl.}$) = (4.058, $-$9.873) with standard deviations along the major axis of the ellipse of 0.845 \masyr\ and along the minor axis of 0.686 \masyr. 
We considered that stars within 4$\sigma$ from the mean values to be M5 member stars. Then we selected our target RGB stars with \vvhbmag\ from our multicolor photometry. 
In Figure~\ref{fig:cmd}, we show our new color-magnitude diagrams (CMDs) for M5 membership stars. The \cnjwl\ CMD shows discrete double RGB sequences \citep[see also][]{lee17, lee19a, lee19b}.
On the other hand, our \chjwl, \nhjwl, and $m1$ CMDs show very broad RGB sequences, due to spreads in the carbon and nitrogen abundances as we will discuss later \citep[also see][]{lee21}.

We also note that M5 contains a very broad range of the horizontal branch (HB) morphology \citep[e.g., see][]{gratton13}. Compared to M3, which is slightly more metal-poor than M5 \citep{lee21}, M5 has a well developed blue HB, most likely due to the presence of the dispersion in its helium abundance as we will discuss later \citep[see also][]{lee17}.

\begin{deluxetable}{lcr}[t]
\tablenum{1}
\tablecaption{Mean [Fe/H] values.\label{tab:feh}}
\tablewidth{0pc}
\tablehead{
\multicolumn{1}{c}{} &
\multicolumn{1}{c}{[Fe/H]} &
\multicolumn{1}{c}{No.} 
} 
\startdata
All   & $-$1.295 $\pm$ 0.039 & 842 \\
\cnw  & $-$1.290 $\pm$ 0.047 & 242 \\
\cns  & $-$1.296 $\pm$ 0.043 & 600 \\
\cnsi & $-$1.293 $\pm$ 0.032 & 477 \\
\cnse & $-$1.323 $\pm$ 0.036 & 124 \\
\hline
\citet{ivans01}    & $-$1.29 $\pm$ 0.05 & 19 \\
\citet{carretta09} & $-$1.35 $\pm$ 0.02 & 136 \\
\citet{gratton13}  & $-$1.33 $\pm$ 0.02 & 30 \\
\citet{husser20}   & $-$1.16 $\pm$ 0.20 & 863 \\
\enddata 
\end{deluxetable}

\begin{deluxetable}{lcccccc}[b]
\tablenum{2}
\tablecaption{Dispersions in Carbon Abundances of the Faint RGBs\label{tab:sigma}}
\tablewidth{0pc}
\tablehead{
\multicolumn{2}{c}{} &
\multicolumn{2}{c}{M5} &
\multicolumn{1}{c}{} &
\multicolumn{2}{c}{M3} \\
\cline{3-4}\cline{6-7}
\multicolumn{2}{c}{} &
\multicolumn{1}{c}{\cnw} &
\multicolumn{1}{c}{\cns} &
\multicolumn{1}{c}{} &
\multicolumn{1}{c}{\cnw} &
\multicolumn{1}{c}{\cns} 
}
\startdata
$\sigma$[C/Fe] && 0.074 & 0.136 && 0.068 & 0.083 \\
\enddata 
\end{deluxetable}

\begin{figure}
\epsscale{1.1}
\figurenum{2}
\plotone{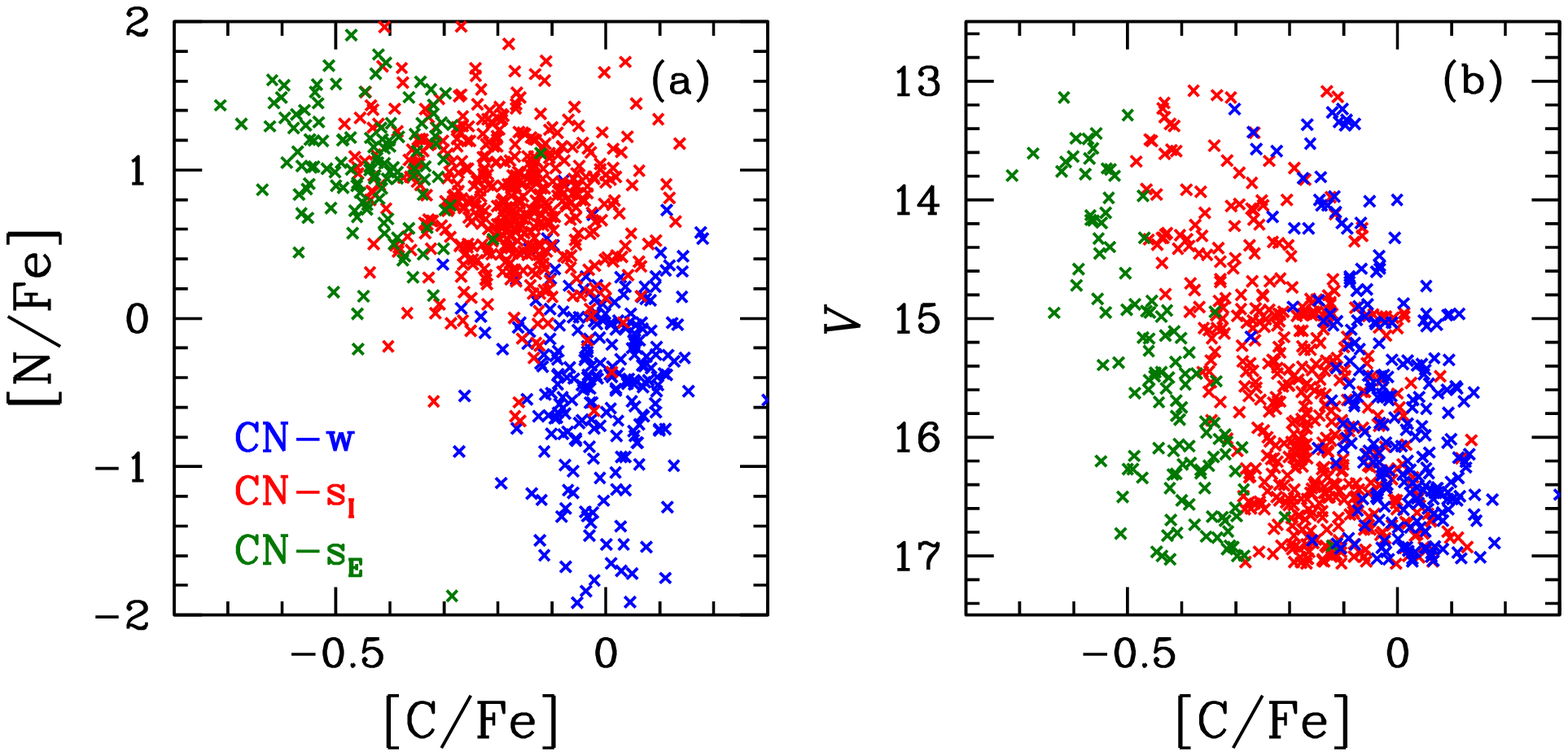}
\caption{
(a) A plot of \cfe\ vs.\ \nfe\ for well measured [\shk\ $\leq$ 0.01 mag] M5 RGB stars with \vvhbmag. The blue color denotes the \cnw\ population, while the red and dark-green colors denote the \cns-I and \cns-E populations. (b) A plot of [C/Fe] vs.\ $V$ magnitude, showing that the boundary between the \cnsi\ and \cnse\ is not set at a fixed [C/Fe] but reflects the variation of carbon abundance against the $V$ magnitude due to the internal mixing during the evolution of low-mass stars.
}\label{fig:cnr}
\end{figure}

\subsection{Elemental Abundances and Populational Tagging}\label{s:abund}
In our previous study of M5, we performed populational tagging based on the \cnjwl\ distribution of RGB stars, showing a discrete bimodal distribution \citep{lee17,lee19a,lee19b}. Here, we take a similar path that will reveal three different populations in M5.

Using the populational tagging for the \cnw\ and \cns\ populations\footnote{The \cnw\ and \cns\ populations of our study are defined to be groups of RGB stars with weak CN band strengths at \cnwave\ (i.e., carbon-normal and nitrogen-normal) and strong CN band strengths (i.e., carbon-poor and nitrogen-rich), respectively, following the conventional nomenclature \citep[e.g.,][]{norris81}.} from our previous study \citep{lee17,lee19a,lee19b}, we derived photometric elemental abundances from our color indices using the similar method that we developed in our previous work \citep{lee21}. 
We obtained the Dartmouth model isochrones for [Fe/H] = $-$1.5, $-$1.4, $-$1.3, $-$1.2, and $-$1.1 dex with [$\alpha$/Fe] = +0.4 dex, and the age of 12.5 Gyr \citep{dartmouth}. We interpolated the effective temperatures and surface gravities from $M_V$ = 3.5 to $-$2.5 mag with a magnitude step size of $\Delta M_V$ = 0.2 mag. Using these stellar parameters, we constructed a series of synthetic spectra with varying elemental abundances with abundance step sizes of [X/Fe] = 0.2 dex.
For our synthetic spectrum calculations, we used the 2011 version of the local thermodynamic equilibrium (LTE) line analysis code MOOG that includes Rayleigh scattering from neutral hydrogen \citep{moog,moogscat} and the atomic/molecular line lists generated from the $linemake$ facility.\footnote{Available at \url{https://github.com/vmplacco/linemake}.}
In total, we calculated more than 210,000 synthetic spectra for our current work. 
Finally, individual synthetic spectra were convolved with our filter transmission functions to be converted to our photometric system.

The photometric metallicity of individual RGB stars can be calculated using the following relation \citep[also see Appendices of][]{lee21},
\begin{eqnarray}
{\rm [Fe/H]} &\approx& f_1(hk_{\rm JWL},~ M_V),\label{eq:FeH}
\end{eqnarray}
and our results are given in Table~\ref{tab:feh}. 
We obtained \feh\ = $-$1.295 $\pm$ 0.039 $\pm$ 0.001 dex (the errors are for the standard deviation and the standard error) and our result appears to be consistent with those from other researchers listed in Table~\ref{tab:feh} to within measurement uncertainties.
Note that we used only those RGB stars with low measurement errors, \shk\ $\leq$ 0.01 mag, which will ensure that our photometric metallicity of individual stars is not affected by measurement error. 
Also importantly, the proper motion study of Gaia EDR3 is not complete in the central part of the cluster. Consequently, our RGB sample with photometric elemental abundances is not a complete sample and tends to be biased toward the outer part of cluster, where the degree of crowding effect from the nearby stars is less severe. In total, we measured metallicity for 242 \cnw\ and 600 \cns\ RGB stars.
As shown in the table, our photometric [Fe/H] values of each population are in excellent agreement with each other, suggesting that M5 is a mono-metallic GC to within $\sigma$\feh\ $\lesssim$ 0.05 dex.

The photometric carbon and nitrogen abundances can be estimated using the above 842 RGB stars as follows,
\begin{eqnarray}
{\rm [C/Fe]} &\approx& f_2(ch_{\rm JWL},
~{\rm [Fe/H]},~ M_V), \label{eq:CFe} \\
{\rm [N/Fe]} &\approx& f_3(nh_{\rm JWL},~ {\rm [Fe/H]},~ M_V).\label{eq:NFe}
\end{eqnarray}
In Figure~\ref{fig:cnr}, we show a plot of [C/Fe] versus [N/Fe] of the M5 RGB stars with \vvhbmag, showing a strong carbon--nitrogen anticorrelation, which is a natural consequence of the CN-cycle hydrogen burning.

It should be noted that the carbon abundance spread in the M5 \cns\ RGB stars is very large. In Table~\ref{tab:sigma}, we compare the \scfe\ of M5 and M3 RGB stars fainter than their RGB bumps (RGBBs). It is believed that RGB stars fainter than the RGBB maintain their initial \cfe\ and \nfe\ abundances and are not affected by the CN cycle accompanied by a noncanonical thermohaline deep mixing that can significantly alter the surface \cfe\ and \nfe\ during the evolution of low-mass stars \citep{charbonnel07}. We note that the \scfe\ of the \cnw\ and \cns\ populations in M3 are comparable to each other \citep{lee21}.
In sharp contrast, the \scfe\ the M5 \cns\ stars is about twice as large as that of the M5 \cnw\ population, indicating that the M5 \cns\ population may contain multiple subpopulations.\footnote{See Figure~5(a) of \citet{lee21} and Figure~9 of \citet{lee19b} for the \pchjwl\ extents of the \cns\ populations in M3 and M5, where
\begin{equation}
\parallel ch_{\rm JWL} \equiv \frac{ch_{\rm JWL} - ch_{\rm JWL,red}}
{ch_{\rm JWL,red}-ch_{\rm JWL,blue}},
\end{equation}
and $ch_{\rm JWL,blue}$ and $ch_{\rm JWL,red}$ are the fiducials of the red and the blue sequences of the \chjwl\ index, respectively. One can find that the extent of the \pchjwl\ of the M5 \cns\ population is significantly large compared to that of the M3 \cns.}
Note that the previous spectroscopic study of M5 by \citet{carretta09} showed an extended Na-O anticorrelation. They proposed that M5 may contain three different populations, the primordial, the intermediate, and the extreme components, based on the location on the plot of [O/Fe] versus [Na/Fe].\footnote{Note that \citet{carretta09} separated three populations in M5 from fixed [O/Fe] and [Na/Fe] values, without considering elemental abundance variations against luminosity, due to the existence of evolutionary effects, such as an internal deep mixing, as \citet{lee10} showed. Therefore, their primordial, intermediate, and extreme components are not exactly the same as our \cnw, \cnsi, and \cnse\ populations \citep[see also][]{lee17}. }

In our previous study of the cluster \citep{lee17,lee19a,lee19b}, we studied the MPs of M5 RGB stars based on the \cnjwl\ distribution, finding two distinctive populations, the \cnw\ and \cns.
Careful examination of the \pchjwl\ distribution of the M5 \cns\ population may suggest that it can be well described by a bimodal distribution. In order to distinguish the \cns\ subpopulations, we applied the expectation maximization (EM) algorithm for the two-component Gaussian mixture model, finding the number ratio of \nrgbsub\ = 79:21 ($\pm$3). 
As shown in Figure~\ref{fig:cnr}, our \cnsi\ and \cnse\ are those with the intermediate [C/Fe] and the most carbon depleted populations, respectively, and they are roughly corresponding to the intermediate and extreme components by \citet{carretta09}, respectively. 
Including the \cnw\ population, which is corresponding to the primordial component devised by \citet{carretta09}, the M5 populational number ratio becomes \nrgbthree\ = 29:56:15 ($\pm$3).

\begin{figure}[t]
\epsscale{.8}
\figurenum{3}
\plotone{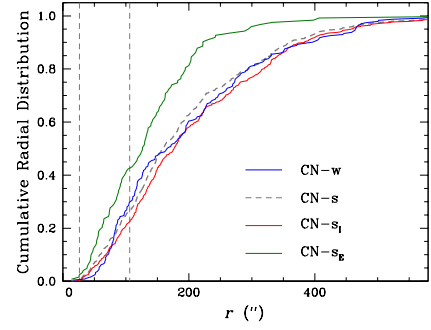}
\caption{
CRDs of the individual populations in M5. Vertical gray dashed lines are for the core and half-light radii. Note that the \cnse\ stars are significantly centrally concentrated than other populations.
}\label{fig:rad}
\end{figure}

\subsection{Cumulative Radial Distributions}
The cumulative radial distributions (CRDs) of MPs in GCs can provide pivotal information on the formation and the dynamical evolution of MPs.
Recent numerical simulations may suggest that the SG with extreme helium abundance formed first in the innermost part of GCs out of gas that experienced proton-capture processes at high temperatures, while the SG with modest helium enhancement formed out of interstellar media diluted from the pristine gas in the more spatially extended region.
The initial formation location of the SG of stars may depend on the physical environment that proto-GCs resided  \citep{calura19}. 
Also, the dynamical effects on smaller stellar masses of the helium-enhanced populations can alter initial CRDs during the course of GC evolution, although the tidal field of the host galaxy could be a major component in determining the relative CRDs of MPs \citep[e.g., see][]{fare18}

In our previous study, we showed that CRDs of the M5 \cnw\ and \cns\ populations are statistically identical up to more than five half-light radius. 
Here we repeated similar tasks for the three populations with low measurement uncertainties and we show our results in Figure~\ref{fig:rad} and Table~\ref{tab:ks}.
The figure clearly shows that the \cnw\ CRD is in good agreement with that of the \cns\ [= \cnsi\ + \cnse].
We performed Kolmogorov--Smirnov (K-S) and Anderson--Darling (A-D) tests to see if they are statistically similar distributions. It is a well-known fact that the K-S test can be sensitively dependent on the near center of the distribution and less dependent on the edges of the distribution, while the A-D test is known to be less vulnerable to such problems. Our results show that the \cnw\ and \cns\ are most likely drawn from the identical parent distributions.
Since the \cnsi\ is the major component of the \cns, the CRD of the \cnsi\ would be similar to that of the \cns.
On the other hand, the CRD of \cnse\ may tell a different story.
The figure clearly shows that the \cnse\ is the most centrally concentrated. Our K-S tests suggest that the CRD of the \cnse\ is significantly different from the others.
Therefore, our results strongly suggest that the \cnse\ population must have formed in the innermost region of M5.

\begin{deluxetable}{ccccc}[t]
\tablenum{3}
\tablecaption{$p$ Values Returned from the K-S and A-D Tests for Individual CRDs\label{tab:ks}}
\tablewidth{0pc}
\tablehead{
\multicolumn{3}{c}{Populations} &
\multicolumn{1}{c}{K-S} &
\multicolumn{1}{c}{A-D} 
}
\startdata
\cnw  & vs.\ & \cns       &  0.505   & 0.211 \\
\cns  & vs.\ & \cnsi      &  0.361   & 0.098   \\
\cns  & vs.\ & \cnse      &  1.05$\times10^{-4}$ & 2.18$\times10^{-6}$ \\
\cnw  & vs.\ & \cnsi      &  0.099 & 0.147  \\
\cnw  & vs.\ & \cnse      &  1.35$\times10^{-5}$ & 1.11$\times10^{-6}$ \\
\cnsi & vs.\ & \cnse      &  6.51$\times10^{-7}$ & 1.49$\times10^{-9}$ \\
\enddata 
\end{deluxetable}

\begin{figure}[t]
\epsscale{1.1}
\figurenum{4}
\plotone{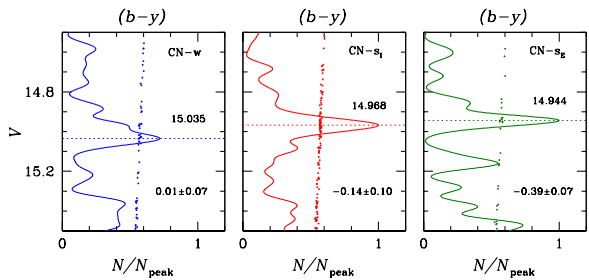}
\caption{
Plots of $(b-y)$ vs. $V$ CMDs and generalized differential LFs of the \cnw, \cnsi, and \cnse\ RGB stars. We show the RGBB $V$ magnitudes of individual populations with gray dotted lines. The numbers in the bottom of each panel denote the mean [C/Fe] values of the RGB stars fainter than the RGBB.
}\label{fig:bump}
\end{figure}

\subsection{Red Giant Branch Bump Magnitudes}
Helium is the second most abundant element but it cannot be detected directly in GC RGB stars due to the lack of any measurable spectral lines.\footnote{For example, see \S6.3.1 of \citet[][and references therein]{lee19b} for the spectroscopic helium abundance measurements using the photospheric \ion{He}{1} $\lambda$5876\AA\ line and chromospheric \ion{He}{1} $\lambda$10830\AA\ line.}
Instead, indirect methods, such as the RGBB magnitudes of a single population, can play an important role to probe the helium contents in GCs.
During the evolution of low-mass stars, RGB stars experience slower evolution and temporary drop in luminosity when the very thin H-burning shell crosses the discontinuity in the chemical composition and lowered mean molecular weight left by the deepest penetration of the convective envelope during the ascent of the RGB, the so-called RGBB \citep[e.g., see][]{renzini88}. 
It is well understood that, at a given age, the RGBB $V$ magnitude decreases with metallicity and increases with helium abundance.
In the study of MPs in a GC, very accurate differential photometry can be attained and, therefore, one can accurately estimate relative helium contents among MPs in a mono-metallic GC \citep{lee15, lee17, lee18, milone18, lee21}.

We derived the generalized differential luminosity functions (LFs) and the RGBB $V$ magnitudes for individual populations in M5 and we show our results in Figure~\ref{fig:bump}. 
We obtained the RGBB $V$ magnitudes of 15.034, 14.962, 14.968, and 14.944 ($\pm$0.030) mag for the \cnw, \cns, \cnsi, and \cnse, respectively. Note that our new measurements for the limited number of \cnw\ and \cns\ stars are consistent with our previous results, 15.038 and 14.970 ($\pm$0.030), respectively \citep{lee17}. Our photometric metallicities of the three populations are in excellent agreement and, therefore, the difference in helium abundance is most likely responsible for the RGBB magnitude difference. As we argued previously \citep{lee17}, our results suggest that the \cnsi\ is likely enhanced in helium by \dy\ $\approx$ 0.026 $\pm$ 0.017 with respect to the \cnw.
The RGBB $V$ magnitude of the \cnse\ population is the brightest and it could be the most helium-enhanced population by \dy\ $\approx$ 0.036 $\pm$ 0.017 with respect to the \cnw\ population.
Our inferred helium abundances are in good agreement with that of \citet{dantona08}, who estimated the helium enhancement of the M5 SG by \dy\ $\approx$ 0.02 -- 0.07 with respect to the FG from their synthetic HB models.

\section{Summary}
With our new photometric indices and theoretical fine model grids for various stellar parameters and abundances using synthetic spectra, we derived \feh, \cfe, and \nfe, the key elements in the GC MP study, of individual RGB stars in M5. Our \feh\ measurements suggest that M5 is a mono-metallic GCs with a small metallicity dispersion, \feh\ = $-$1.295 $\pm$ 0.039 $\pm$ 0.001 dex .

We showed that the dispersion in \cfe\  of the M5 \cns\ is very large compared to those of the M5 \cnw\ or the M3 \cnw\ and \cns, suggesting that the M5 \cns\ is composed of MPs.
Our new analysis revealed that the M5 \cns\ can be well described by at least two populations, the \cnsi\ and \cnse, and, as a consequence, M5 contains at least three distinctive MPs, including the \cnw.  We obtained the number ratio of \nrgbthree\ = 29:56:15 ($\pm$3).
The \cnse\ is significantly more centrally concentrated than the others, while the \cns\ and \cnsi\ have the similar CRDs as the \cnw\ does, consistent with our previous result \citep{lee17}.

We found a correlation between the mean \cfe\ and the RGBB $V$ magnitude in individual populations in M5, in the sense that the RGBB $V$ magnitude increases with the carbon abundance. Since the three populations in M5 have almost identical metallicity, the difference in the RGBB $V$ magnitude can be interpreted as the difference in helium abundance. We estimated that the \cnsi\ and \cnse\ populations are enhanced in helium by \dy\ $\approx$ 0.026 $\pm$ 0.017 and 0.036 $\pm$ 0.017, respectively, with respect to the \cnw\ population, which is thought to have a normal helium abundance for its metallicity.

Our results for the CRDs and inferred helium abundances can be nicely explained by recent numerical simulations by others \citep[e.g.,][]{fare18, calura19}. After the termination of the \cnw\ supernovae explosions, the \cnse\ (i.e., the most helium-enhanced SG by \citealt{calura19}) formed out of gas that experienced proton-capture processes at high temperatures, rich in helium and nitrogen and poor in carbon, in the innermost region. Shortly after, the \cnsi\
(i.e., the modest helium-enhanced SG by \citealt{calura19}) formed out of gas diluted from the pristine gas in the more spatially extended region.
It is likely that the proto-GC of M5 resided in a dense density external gas environment where the modest helium-enhanced population will be the major component \citep{calura19}, consistent with our new result for M5. 

The different degree of diffusion processes due to slightly smaller stellar masses with helium abundance cannot be completely ruled out to explain the similar CRDs between the \cnw\ and \cnsi.
However, as \citet{fare18} discussed, the Milky Way tidal field would be a more important factor to determine the CRDs of M5, since it has a very elongated orbit in our Galaxy, with the perigalacticon distance of 2.90 $\pm$ 0.05 kpc and the apgalacticon distance of 24.20 $\pm$ 1.00 \citep{baumgardt19}. 
At the same time, M5 must have experienced a significant degree of shock-induced mass-loss when it passed through the Galactic disk and bulge \citep[e.g.,][]{gnedin97}, that may cause the truncation in the radial distributions of the \cnw\ and \cnsi, resulting in similar CRDs between the two \citep[e.g., see][]{lee18}. 

\acknowledgements
J.-W.L.\ thanks an anonymous referee for a careful review of the paper and many helpful suggestions.
He acknowledges financial support from the Basic Science Research Program (grant No.\ 2019R1A2C2086290) through the National Research Foundation of Korea (NRF) and from the faculty research fund of Sejong University in 2019.

\end{document}